\documentclass[twocolumn]{jpsj2}
\usepackage{epsfig}

\title{Real Space Effective Interaction 
and Phase Transition in the Lowest Landau Level}
\author{Naokazu {\sc Shibata} and Daijiro {\sc Yoshioka}}
\inst{
Department of Basic Science, University of Tokyo,
Komaba 3-8-1 Meguro-ku, Tokyo 153-8902, Japan
}
\recdate{March 16, 2004}

\abst{
The transition between the stripe state and the liquid
state in a high magnetic field is studied by the
density-matrix renormalization-group (DMRG) method.
Systematic analysis on the ground state of two-dimensional 
electrons in the lowest Landau level shows that the transition from the 
stripe state to the liquid state at $\nu\sim 3/8$ is 
caused by a reduction of repulsive 
interaction around $r\sim 3\ell$. 
The same reduction of the interaction also stabilizes
the incompressible liquid states at $\nu=1/3$ and $2/5$,
which shows a similarity between the two liquid states
at $\nu\sim 3/8$ and $1/3$. 
It is also shown that 
the strong short-range interaction around $r\sim \ell$
in the lowest Landau level makes qualitatively different
stripe correlations compared with that in higher Landau levels.
}

\kword{stripe, liquid, transition, fractional quantum Hall effect, 
two dimension, density matrix, renormalization group}

\def\runauthor{Naokazu {\sc Shibata} and Daijiro {\sc Yoshioka}}
\begin{document}

\maketitle

\section{Introduction} 
In two-dimensional systems under a high perpendicular magnetic field, 
the kinetic energy of electrons is completely quenched,
and macroscopic degeneracy appears in each Landau level. 
The macroscopic degeneracy is lifted by Coulomb interaction
and various interesting ground states, including 
incompressible liquids,\cite{Lagh,Yoshi} compressible liquids,\cite{Jain,Halp} 
CDW states\cite{Kou1,Kou2} called stripes, bubbles 
and Wigner crystal are realized
depending on the filling $\nu$ of Landau levels. 

In the lowest Landau level, the incompressible liquid states
are realized at various fractional fillings $\nu=n/(2n\pm 1)$,
which are known as fractional quantum Hall 
states.\cite{FQHE1,FQHE2}
The Wigner crystal is realized at low fillings
and its formation has been observed  as 
a transition to an insulating state below 
$\nu\sim 1/5$, which is thought to be caused by the pinning 
of Wigner crystal by impurity potentials.\cite{WC1,WC2,WC3}

Concerning the ground state between the incompressible liquid 
states at $\nu=n/(2n\pm 1)$ and $1/5$,
there still exist many questions. Even though
recent numerical calculations based on the density-matrix 
renormalization-group (DMRG) method show the existence of weak 
stripe states,\cite{Shib1} clear experimental evidence of the 
stripe formation has not yet been obtained.
The stripe state obtained in the DMRG has
significantly small amplitude of the stripes and the 
short-range correlations
are similar to those of the Wigner crystal.\cite{Shib1} 
It has also been shown that a transition from the stripe state
to a liquid state occurs as the 
short-range repulsion between the electrons is reduced,\cite{stripe} 
which means that the stripe state in the lowest Landau level
is realized only in narrow quantum wells.

In this paper we investigate the ground state at $\nu= 3/8$, 
where the stripe state is realized in ideal two-dimensional 
systems,\cite{Shib1,stripe}
and show why the stripe state in the lowest Landau level
is realized only in narrow quantum wells.
From the analysis on the change in the pair correlation function
through the transition to a liquid state,
we show that the stripe state is stabilized by the 
strong repulsive interaction around $r\sim 3\ell$.
We also show that the same repulsive interaction around 
$r\sim 3\ell$ reduces the excitation gap of 
the incompressible liquid states at $\nu=1/3$ and $2/5$, 
which suggests a similarity between the incompressible 
liquid state at $\nu=1/3$ and the liquid state at $\nu=3/8$.
We finally explain the origin of the difference in the stripe 
correlation between the lowest and higher Landau levels
by comparing the Coulomb interaction projected onto each 
Landau level.

\section{Model and Method} 
We use the Hamiltonian of two-dimensional 
electrons in a perpendicular magnetic field. 
Since the kinetic energy of the electrons is completely 
quenched, we can omit this energy.
The Hamiltonian of the electrons in the lowest Landau level
is then written by
\begin{equation}
H=\sum_{i<j} \sum_{\bf q} e^{-q^2 \ell^2/2} V(q) \ 
e^{i{\bf q} \cdot ({\bf R}_i-{\bf R}_j)} 
\label{Coulomb}
\end{equation}
where ${\bf R}_i$ is the guiding center coordinate of the $i$th
electron, which satisfies the commutation relation,
$[{R}_{j}^x,{R}_{k}^y]=i\ell^2\delta_{jk}$
and $V(q)$ is the Fourier transform of the
Coulomb interaction, which is given by 
$2\pi e^2/\varepsilon q$
in ideal two-dimensional systems. 
We set the magnetic length $\ell$ equal to unity,
and use $e^2/\varepsilon \ell$ as units of energy.
We assume that the magnetic field is strong enough to 
polarize the spins and suppress the Landau level mixing.

In order to obtain the ground-state wave function, we use the
DMRG method, which was originally developed for one-dimensional
quantum systems.\cite{DMRG}
We apply this method to two-dimensional systems divided into 
unit cells $L_x\times L_y$ with the periodic boundary 
conditions for both $x$- and $y$-directions.\cite{Shib,DY,Shib3} 
This method enables us to obtain the essentially exact 
ground state of large systems extending the limitation of
the exact diagonalization method with controlled accuracy.
The truncation error in the ground-state wave function
is typically $10^{-4}$ with keeping 200 eigenvectors 
of the density matrix.

We calculate the ground-state energy and the wave function for 
various size of systems with up to 18 electrons 
in the unit cell, and analyze the pair correlation function 
$g({\bf r})$ defined by
\begin{equation}
g({\bf r}) \equiv \frac{L_x L_y}{N_e(N_e-1)}\langle 
\Psi | \sum_{i\neq j} \delta({\bf r}+{\bf R}_i-{\bf R}_j)|\Psi
\rangle
\end{equation}
where $|\Psi\rangle$ is the ground state with $N_e$ being the 
number of electrons in the unit cell. 
Since the period of the stripes is artificially modified 
by changing the aspect ratio $L_x/L_y$, we need to find 
the unit cell that has the energy minimum with respect to $L_x/L_y$.
The correlation functions in such unit cell
are expected to have the correct period of the stripes 
realized in the thermodynamic limit.

\begin{figure}[b]
\begin{center}\leavevmode
\epsfxsize=80mm \epsffile{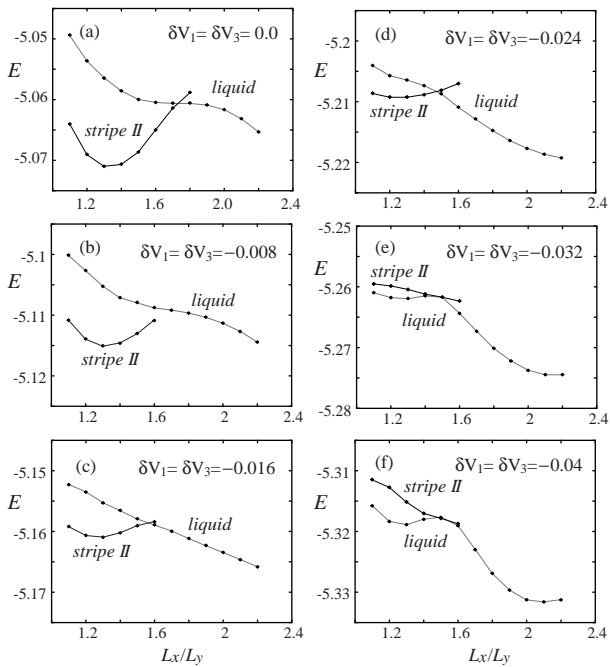}
\caption{The energies of the stripe state and the liquid state
at $\nu=3/8$. $N_e=12$.
The stripe state is not realized around $L_x/L_y\sim 2$.
{\it stripe-II} represents the stripe state in the lowest 
Landau level.
}
\end{center}\end{figure}

\section{Pseudopotentials and Phase Transition} 
In ideal two-dimensional systems, the stripe ground state 
in the lowest Landau level is realized around $\nu=0.42$, 
$0.37$ and between $0.32$ and $0.15$.\cite{Shib1} 
However, experimental evidence of the stripe formation
has not been obtained. Instead, the transport experiments
on ultra high mobility wide quantum wells\cite{WQW}
suggest the formation of liquid states 
even at the fillings where the stripe state 
is numerically obtained.  
Since the finite width of the two-dimensional system
reduces the short-range repulsion between the electrons,
we here study the stability of the stripe state
against the reduction of short-range interaction.

\begin{figure}[t]
\begin{center}\leavevmode
\epsfxsize=85mm \epsffile{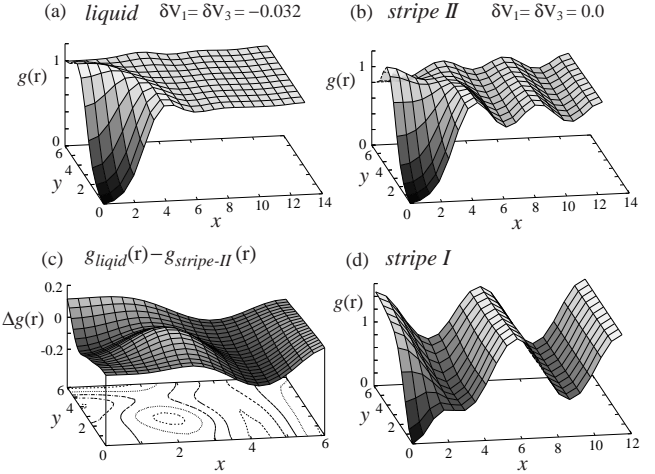}
\caption{The pair correlation functions $g({\bf r})$.
$N_e=18$. 
(a) The liquid state at $\nu=3/8$ for
$\delta V_1=\delta V_3=-0.024$.
(b) The stripe state at $\nu=3/8$ for $\delta V_1=\delta V_3=0$ 
({\it stripe-II}).
(c) The difference in $g({\bf r})$ shown in (a) and (b).
(d) The stripe state in higher Landau level of $N=2$ at $\nu_N=3/7$
in guiding center coordinates ({\it stripe-I}).
}\label{fig2}
\end{center}\end{figure}

We first consider the effect of the decrease in
short-range components of Haldane's 
pseudopotentials,\cite{pseudo} $V_1$ and $V_3$.
Since the change in $V_1$ does not affect 
the relative energy difference between the stripe and liquid 
states,\cite{stripe} we consider the case of 
$\delta V_1=\delta V_3$, where 
$\delta V_m = V_m-(V_m)_{\mbox{\scriptsize 2D}}$ with
$(V_m)_{\mbox{\scriptsize 2D}}$ being the 
pseudopotentials of the pure Coulomb interaction in 
ideal two-dimensional systems.
We calculate the ground state for wide 
range of $\delta V_3$ to confirm the transition to the liquid state.
The ground-state energies of the stripe state and the liquid state 
at $\nu=3/8$ are shown in Fig.~1.
We find relative decrease in the energy of the
liquid state with the decrease 
in $\delta V_3$. 
For small $V_3$ below $\delta V_3\sim -0.032$,
the liquid state has lower energy independent of $L_x/L_y$,
which clearly shows the existence of a transition to the
liquid state.

The correlation functions for the liquid ground state
at $\delta V_1=\delta V_3=-0.024$ and 
the stripe ground state at
$\delta V_1=\delta V_3=0$
are presented in Figs.~2 (a) and (b), respectively. 
The size dependence of the stripe correlation 
has been analyzed for systems 
with up to 24 electrons in the unit cell,
and the stripe correlations
are shown to be almost the same even for large systems.\cite{stripe}
Since the correlation function and the total momentum
of the stripe and liquid states are both different,
we expect a first order transition 
even in the thermodynamic limit.
The critical value of $\delta V_3$ 
in the case of $\delta V_m = 0$ for $m\ge 5$
is expected to be around $-0.02$, below which
the energy of the liquid state has clear minima
at $L_x/L_y\sim 1.3$ and $2.2$, which continuously 
connect to the corresponding minima 
of the liquid state below $\delta V_3\sim -0.032$.

To further investigate the nature of the transition,
we next compare the correlation functions between the two states.
The difference between the two
correlation functions in Figs.~2 (a) and (b)
is presented in Fig.~2 (c). 
We find a clear peak at $x\sim y\sim 2$ and a dip around $r\sim 4$,
which means that the electrons in the liquid state are more likely to
approach each other.
The existence of the clear peak at
$r\sim 3$ $(x\sim y\sim 2)$ also suggests that the 
energy difference between the two states is
sensitive to the change in the interaction at $r \sim 3$.

\begin{figure}[t]
\begin{center}\leavevmode
\epsfxsize=70mm \epsffile{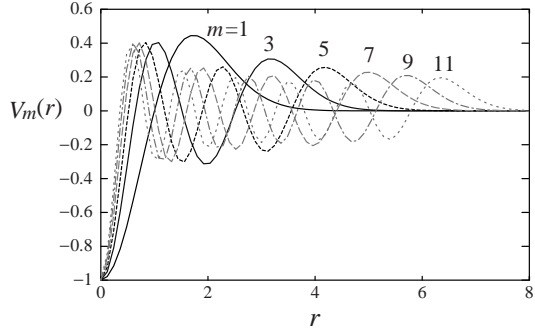}
\caption{The Haldane's pseudopotentials in
real space. $m$ is the relative angular momentum
between the two electrons and only odd integers are 
relevant for single component fermions.
}\label{fig3}
\end{center}\end{figure}

\begin{figure}[t]
\begin{center}\leavevmode
\epsfxsize=80mm \epsffile{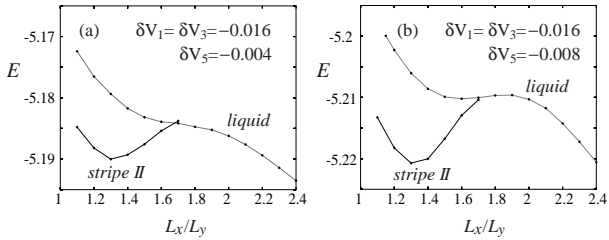}
\caption{
The energies of the stripe state and the liquid state
at $\nu=3/8$.
The number of electrons in the unit cell $N_e$ is 12. 
(a) $\delta V_5=-0.004$. (b) $\delta V_5=-0.008$.
}\label{fig4}
\end{center}\end{figure}

Figure 3 shows the Haldane's pseudopotentials\cite{pseudo} 
for relative momentum $m$
represented in real space which is defined by the following equation,
\begin{equation}
{\cal V}_m({\bf r})=\frac{1}{2\pi}\int d^2q\ L_m(q^2) e^{-q^2/2} 
\exp(-i{\bf q \cdot r}) 
\end{equation}
where $L_m(x)$ are Laguerre polynomials. The effective 
interaction ${\cal V}_{\mbox{\scriptsize eff}}(\bf r)$ 
between the two-dimensional electrons 
in guiding center coordinates is then given by 
\begin{equation}
{\cal V}_{\mbox{\scriptsize eff}}({\bf r})=\sum_m 2 V_m {\cal V}_m({\bf r}).
\end{equation}
We find that the pseudopotential ${\cal V}_m({\bf r})$
for $m=3$ has a maximum at $r\sim 3$,
which means the decrease in $V_3$ stabilizes the 
liquid state.
This is consistent with the results in Fig.~1.

In contrast to the pseudopotential for $m=3$,
the pseudopotential for $m=5$ has a minimum around $r\sim 3$,
which suggests a relative increase in the energy of the 
liquid state with the decrease in $V_5$.
This is actually shown in Fig.~4,
where we find clear energy minimum for the stripe state,
which shows the stability of the stripe state even in the 
case of $\delta V_1=\delta V_3= -0.016$.

With increasing the relative momentum $m$, the pseudopotential 
${\cal V}_m(r)$ oscillates rapidly as shown in Fig.~3. 
To see the effect of large $m$, we next decrease
$V_m$ up to $m=11$. The results for the case of 
$\delta V_5=\delta V_7=\delta V_9=\delta V_{11}=\delta V_3/4$
is presented in Fig.~5.
We find the energy of the liquid state relatively decreases 
with the decrease in $V_m$.
This means the effect of $V_5$ 
is canceled by the decrease in $V_m$ of higher $m$.
Indeed we find a peak of pseudopotential at $r\sim 3$ 
for $m=7$.
The tendency of the stabilization of the liquid state 
does not change even for larger systems 
of 18 electrons in the unit cell as shown in Figs.~5 (c) and (d).

The above results show that the transition to the liquid state 
is mainly controlled by $V_3$, whose effective interaction
has a maximum at $r\sim 3$.
The stabilization of the liquid state is caused by 
the decrease in the effective repulsion around $r\sim 3$
and this decrease makes the electrons to approach each other
with the enhancement in quantum fluctuations 
through the overlap of their wave functions.

\begin{figure}[t]
\begin{center}\leavevmode
\epsfxsize=80mm \epsffile{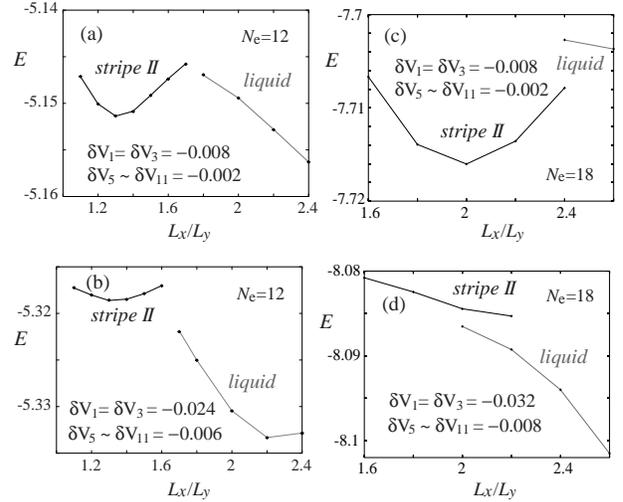}
\caption{
The energies of the stripe state and the liquid state at $\nu=3/8$.
$\delta V_1=\delta V_3$ and $\delta V_5=\delta V_7=\delta 
V_9=\delta V_{11}=\delta V_3/4$.
$N_e=12$ in (a) and (b). $N_e=18$ in (c) and (d).
}\label{fig5}
\end{center}\end{figure}

\section{Similarity to the Incompressible Liquids}

To clarify the similarity and the difference between the
two liquid states at $\nu\sim 3/8$ and $\nu=n/(2n+1)$,
we here consider the role of short-range repulsion
in the incompressible states at $\nu=n/(2n+1)$.
At $\nu=1/3$, it has been shown that the Laughlin state 
is an exact ground state in the limit of 
$V_1 \rightarrow \infty$.\cite{Lagh} 
This means the incompressible 
liquid state is stabilized by the increase in $V_1$.
Figure 6 actually shows that the excitation gap at $\nu=1/3$ and $2/5$
monotonically decreases with the decrease in $V_1$ and 
it seems to vanish at $\delta V_1 \sim -0.1$ \cite{pseudo}
whose value is 
comparable to the size of the gap at $\nu=1/3$ at $\delta V_m =0$. 
This is contrasted to the fact that 
$V_1$ does not make clear energy difference between
the liquid and stripe states at $\nu\sim 3/8$.\cite{stripe}

\begin{figure}[t]
\begin{center}\leavevmode
\epsfxsize=70mm \epsffile{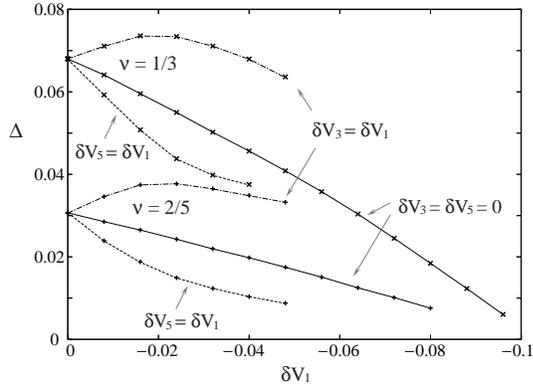}
\caption{
$V_1$, $V_3$ and $V_5$ dependences of the excitation gap 
$\Delta$ in the incompressible
liquid states at $\nu=1/3$ and $2/5$. $N_e=12$.
}\label{fig6}
\end{center}\end{figure}

We next see the effect of $V_3$. 
Figure 6 also shows the gap obtained in the case of 
$\delta V_3=\delta V_1$.
We find the gap is always larger 
than that for $\delta V_3=0$, which means the reduction in $V_3$
enhances the excitation gap.
This is similar to the stabilization of the liquid state 
at $\nu\sim 3/8$ with the decrease in $V_3$.
To further confirm the similarity to the liquid state at
$\nu\sim 3/8$, we next decrease $V_5$. The gap at 
$\delta V_5=\delta V_1$  
is presented in Fig.~6 with the dotted line. In this case
the gap is smaller than that obtained at $\delta V_5=0$, which
shows the decrease in $V_5$ destabilizes the liquid state. 
These results suggest that the liquid state at $\nu=3/8$
has a similar character to the liquid states at $\nu=n/(2n+1)$ 
although the liquid states at $\nu=n/(2n+1)$ have large 
excitation gap, which is controlled by $V_1$.

\section{Effective Potential in Higher Landau Levels}
We finally consider the origin of the difference
in the stripe correlation between the lowest and higher 
Landau levels.
The stripe state in the lowest Landau level is 
qualitatively different from the stripe state 
in higher Landau levels as shown in Figs.~2 (b) and (d). 
The stripes in higher Landau levels 
have large amplitude and the pair correlation function
has a shoulder structure around $r\sim 2$ along the perpendicular
direction to the stripes. 
These results show that the electrons are likely to form 
clusters in higher Landau levels in guiding center coordinates.

\begin{figure}[t]
\begin{center}\leavevmode
\epsfxsize=73mm \epsffile{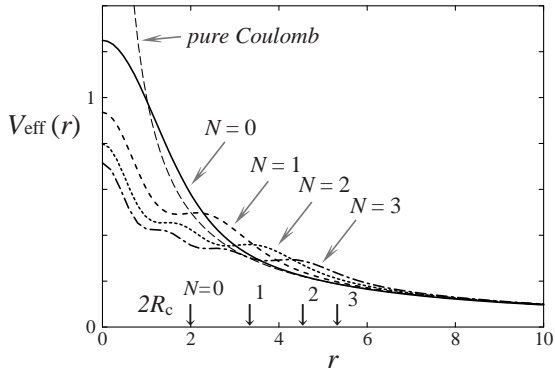}
\caption{
The effective interaction between the electrons
in the $N$th Landau level in guiding center coordinates.
$R_c$ is the classical cyclotron radius.
}\label{fig7}
\end{center}\end{figure}

\begin{figure}[t]
\begin{center}\leavevmode
\epsfxsize=75mm \epsffile{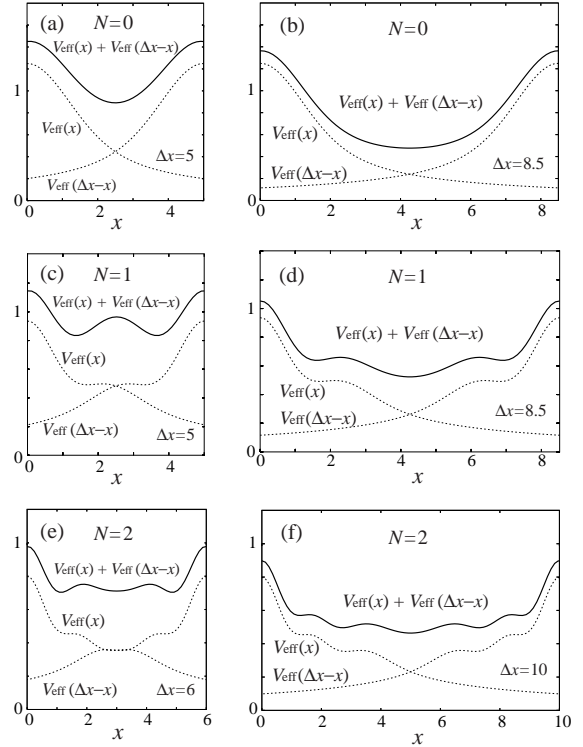}
\caption{
The effective potentials in the $N$th Landau level 
made by two electrons fixed at $x=0$ and $x=\Delta x$
in guiding center coordinates. $\Delta y=0$.
}\label{fig8}
\end{center}\end{figure}

The ground state in each Landau level is determined by the 
Coulomb interaction projected onto each Landau level.
This interaction is represented by a set of
Haldane's pseudopotentials $V_m$.
In Fig.~7 we show the real space representation 
of the pseudopotentials defined in eq.~(4).
In the lowest Landau level we find monotonic decrease 
in the pseudopotential, but in higher Landau levels
we find non-monotonic decrease.
The non-monotonic structure appears below  
twice the classical cyclotron radius $2R_c=2\sqrt{2N+1}$, 
which is shown by the arrows in Fig.~7.

When we fix the guiding centers of the two electrons, 
the other electrons feel the potential made by the two fixed electrons,
which is shown in Fig.~8 by the solid lines,
where $\Delta x$ is the distance between the
two fixed electrons in guiding center coordinates. 
In the lowest Landau level, we always find the potential
minimum at the center of the two fixed electrons.
This is due to the monotonic decrease in the effective 
interaction $V_{\mbox{\scriptsize eff}}$. 
Thus the electrons always keep away 
from the other electrons, and this is the reason why the 
short-range correlations
in the type-{\it II} stripe state are similar to those of Wigner crystal.
In higher Landau levels, however,
the effective potential has potential minima near the two fixed electrons,
which are made by the non-monotonic decrease in 
$V_{\mbox{\scriptsize eff}}$.
These potentials suggest that 
the electrons in higher
Landau levels tend to form clusters when
the mean distance between the electrons 
is not so long compared with the classical cyclotron 
radius $R_c$ as is discussed in the existing paper.\cite{Effec} 
This is the origin of the formation
of the stripes and bubbles in higher Landau levels. 
Thus the difference in the stripe correlation 
between the lowest and higher Landau levels
is naturally explained from the qualitative difference in the
effective interaction between the electrons.

\section{Summary}
In the present study we have calculated the ground state
of two-dimensional electrons for various pseudopotentials. 
The obtained results show that the
liquid state at $\nu\sim 3/8$ is stabilized by the 
decrease in the effective repulsion around $r\sim 3$.
Similar stabilization of the liquid state
at $\nu=1/3$ and $2/5$ suggests that
the liquid states at $\nu\sim 3/8$ and $n/(2n+1)$ 
have a similar character although the liquid states 
at $\nu=n/(2n+1)$ have
large excitation gap which is enhanced by the increase in 
$V_1$.
We have also investigated the reason of the difference 
in the stripe correlations between the
lowest and higher Landau levels and shown that
the monotonic real-space profile of the effective interaction 
in the lowest Landau level is the origin of the difference.

\section*{Acknowledgement}
The present work is supported by
Grant-in-Aid No.~15740177 from MEXT Japan and No.~14540294 from JPSJ.

\end{document}